# Oriented Three-Dimensional Magnetic Biskyrmion in MnNiGa Bulk Crystals

*Xiyang Li, Shilei Zhang, Hang Li, Diego Alba Venero, Jonathan S White, Robert Cubitt, Qingzhen Huang, Jie Chen, Lunhua He, Gerrit van der Laan, Wenhong Wang, Thorsten Hesjedal, and Fangwei Wang\**

((Optional Dedication: Dedicated to the memory of Shou-Cheng Zhang))

X. Y. Li, H. Li, Prof. L. H. He, Prof. W. H. Wang, Prof. F. W. Wang
Beijing National Laboratory for Condensed Matter Physics
Institute of Physics
Chinese Academy of Sciences
Beijing 100190, China
E-mail: fwwang@iphy.ac.cn

Dr. S. L. Zhang, Prof. T. Hesjedal
Department of Physics, Clarendon Laboratory
University of Oxford
Oxford OX1 3PU, United Kingdom

Dr. D. Alba Venero
ISIS
STFC Rutherford Appleton Laboratory
Chilton, Didcot OX11 0QX, United Kingdom

Dr. J. S. White
Laboratory for Neutron Scattering and Imaging
Paul Scherrer Institute
CH-5232 Villigen, Switzerland

Dr. R. Cubitt
Institut Laue-Langevin
6 rue Jules Horowitz, 38042, Grenoble, France

Dr. Q. Z. Huang
NIST Center for Neutron Research
National Institute of Standards and Technology
100 Bureau Drive, Gaithersburg, MD 20899, United States

Dr. J. Chen
China Spallation Neutron Source
Dongguan 523808, China

Prof. G. van der Laan
Magnetic Spectroscopy Group
Diamond Light Source
Didcot OX11 0DE, United Kingdom

X. Y. Li, H. Li, Prof. W. H. Wang, Prof. F. W. Wang
School of Physical Sciences
University of Chinese Academy of Sciences

Beijing 101408, China

Prof. L. H. He, Prof. W. H. Wang, Prof. F. W. Wang
Songshan Lake Materials Laboratory
Dongguan 523808, China

A biskyrmion consists of two bound, topologically stable skyrmion spin textures. These coffee-bean-shaped objects have been observed in real-space in thin plates using Lorentz transmission electron microscopy (LTEM). From LTEM imaging alone, it is not clear whether biskyrmions are surface-confined objects, or, analogously to skyrmions in non-centrosymmetric helimagnets, three-dimensional tube-like structures in bulk sample. Here, we investigate the biskyrmion form factor in single- and polycrystalline MnNiGa samples using small angle neutron scattering (SANS). We find that biskyrmions are not long-range ordered, not even in single-crystals. Surprisingly all of the disordered biskyrmions have their in-plane symmetry axis aligned along certain directions, governed by the magnetocrystalline anisotropy. This anisotropic nature of biskyrmions may be further exploited to encode information.

Systems consisting of elastically coupled arrangements of particles or quasiparticles are a common, yet intriguing many-body problem in physics. First, the elementary unit has to be identified, and its effective potential has to be determined. Next, the interactions among these particles will lead to characteristic dynamics of the ensemble. Depending on the nature of the particles, an order parameter can be assigned to describe a phase transition.[1] For example, in a crystallization process, atoms are usually treated as isotropic particles, and the ordering is evaluated by a scalar structure factor. Nevertheless, in many occasions, these atoms carry magnetic moments that are anisotropic, and which contribute an extra degree of freedom. Consequently, an additional order parameter (magnetization) emerges to describe how these moments order, appearing as a vector field.[2] Moreover, the elementary units can have internal degrees of freedom, which have to be taken into account to understand the emergence of order on all length scales.

Magnetic skyrmions are field-like solutions that carry nontrivial topological properties, and which can be regarded as quasiparticles in continuum field theory.[3,4] Discovered in noncentrosymmetric helimagnets[5,6] and later found in magnetic monolayers and multilayers,[7–9] frustrated magnet,[10,11] as well as in polar magnets,[12] skyrmions are axial-symmetric and localized magnetization configurations with different point-group symmetries of either $D_n$, $T$, $O$ (Bloch-type) or $C_{nv}$ (Néel-type).[13] They can be written as $\mathbf{m} = (m_x, m_y, m_z) = (\sin\Theta \cos\Psi, \sin\Theta \sin\Psi, \cos\Theta)$ in polar coordinates $(\rho, \psi)$, where $\Theta = \Theta(\rho)$ is the radial profile function; and $\Psi = \psi + \chi$, where phase $\chi$ defines the helicity.[3,4] These two types of skyrmions nearly always have isotropic inter-skyrmion potentials,[14] as evidenced by the long-range ordered, hexagonal close-packed skyrmion lattice arrangement.[15] Therefore, the ordering of the skyrmions has scalar nature, justifying rigid-body modeling using molecular dynamics approaches.[16] A major modification of this rigid-body model has to be made to include the description of so-called antiskyrmions which are composed of $D_{2d}$ skyrmions.[17]

These antiskyrmions are anisotropic, thus their interaction potential becomes twofold symmetric, breaking $O(2)$ symmetry. The anisotropic nature of antiskyrmions may therefore also influence their lattice order.[17]

Such anisotropic properties are most pronounced in a biskyrmion system in which the elementary quasiparticles are molecule-like bound skyrmion pairs with $C_{2v}$ symmetry, as observed in LTEM experiments in lamella of $La_{2-2x}Sr_{1+2x}Mn_2O_7$ ($x = 0.315$),[18] MnNiGa,[19–21] $Cr_{11}Ge_{19}$,[22] and FeGd amorphous films.[23] In the stereographic projection representation, a biskyrmion molecule can be written as a complex number $\Omega$:[24]

$$\Omega(d, \Psi, \lambda, \eta) = \frac{m_x + i m_y}{1 + m_z} = \Omega_\chi \Omega_{-\chi} \qquad (1)$$

where $\Omega_\chi$ and $\Omega_{-\chi}$ are two axial-symmetric skyrmions with opposite helicity angle, $\chi$ and $-\chi$. The exact biskyrmion structure is further determined by the skyrmion-skyrmion bond distance $d$, the azimuthal rotation angle $\Psi$, the polarity $\lambda$ (the magnetization direction of the core of $\Omega_\chi$), and a shape factor $\eta$. $\eta$ is the ratio between the long and short axes when distorting an otherwise rotation-symmetric skyrmion into an elliptical shape. **Figure 1** illustrates the key parameters that describe a $C_{2v}$ biskyrmion and their corresponding simulated magnetic form factor, respectively. From the illustration, it is clear that the principle $C_2$ axis defines a directional axis of the biskyrmion motif. In other words, unlike for isotropic skyrmions, biskyrmion molecules have an internal degree of freedom where the in-plane direction $\Psi$ can in principle take arbitrary values. Consequently, when considering the assembly of biskyrmions into ordered lattices, the biskyrmion directional order also has to be taken into account, in addition to their relative positions. For example, one can envision a hexagonally ordered biskyrmion lattice, in which the individual biskyrmion axes can orient in different ways, e.g., parallel, antiparallel, or perpendicular to each other.

So far, the experimental characterization of biskyrmions was mainly based on LTEM studies,[18–23] which provided microscopic insight into their real-space order in quasi-two-dimensional thin plate samples. An in-plane magnetization of such a biskyrmion configuration is shown in **Figure 1g**. In most of these materials systems,[18,19,23,25] the energy density functional $\omega$, and the total energy $E$, can be written in the micromagnetic framework as:

$$\omega = A(\nabla \mathbf{m})^2 - K_u m_z^2 - \mathbf{B} \cdot \mathbf{m}$$
$$E = \int \omega d^3 r + E_{\text{dipolar}} \quad (2)$$

where $A$ is the exchange stiffness constant, $K_u$ is the uniaxial anisotropy, $\mathbf{B}$ is the external magnetic field, and $E_{\text{dipolar}}$ is the non-local dipole-dipole interaction energy.

A stripe domain state ($\mathbf{B} = 0$) is observed experimentally and modeling shows its origin to be stemming from the competition between the uniaxial anisotropy and demagnetization energy,[4] with arbitrary directions of the stripe domains, and with different domain forming on large length scales.[25] When applying a magnetic field, biskyrmions nucleate directly from the stripes.[18,19,22,23] Therefore, a biskyrmion lattice can be observed locally, and the lattice orientation is locked, given by the orientation of the stripes from which they evolved. This locking is observed in $La_{2-2x}Sr_{1+2x}Mn_2O_7$ ($x = 0.315$),[18] MnNiGa,[19] and FeGd thin films.[23] In $Cr_{11}Ge_{19}$, in which the Dzyaloshinskii-Moriya interaction also plays a role, the biskyrmion lattice seems to form in the same way.[22]

Nevertheless, even in the absence of an ordered biskyrmion lattice phase, the in-plane direction of the biskyrmions (characterized by azimuthal angle $\Psi$) is aligned along certain directions in MnNiGa,[19–21] independent of the pinch-off direction of the initial stripe phase.[18] However, the physical origin that determines this direction has not been reported. Furthermore, it still remains an open question whether biskyrmions are a purely surface-related phenomenon (as only thin lamella with non-bulk-like properties had been investigated in LTEM),[26,27] or whether biskyrmions extend into the bulk, analogous to regular skyrmions.[28] On the one hand, it seems intuitive to assume that biskyrmions behave like pairs of skyrmions, inheriting their fundamental topological properties. On the other hand, their deeply distorted magnetization distribution could also lead to strongly surface-confined, bobber-like minimum energy objects.[26,27] Here, we perform SANS on bulk crystal MnNiGa (the crystal information shown in Figure 1h, 1i, and 1j) to provide the answers to these two key questions, and which have remained elusive so far.

**Figure 2a** shows the typical SANS geometry used in our experiments and **Figure 2b**, **2f**, **2j**, and **2n** show the Ewald sphere representation for the SANS geometry for different magnetic reciprocal space configurations. First, a polycrystalline sample is measured using the SANS2d instrument at ISIS with $\mathbf{k}_i \perp \mathbf{B}$ geometry to demonstrate the bulk effect of biskyrmions, and the data shown in panels **Figure 2d**, **2h**, **2l**, and **2p**. Then, a single-crystalline sample is measured using the D11 instrument at ILL with $\mathbf{k}_i \parallel \mathbf{B}$ geometry to study the form factor of the biskyrmion. For the single-crystalline sample, the incident neutron beam along $\mathbf{k}_i$ is parallel to $z$ (Figure 2a), while the magnetic field is applied along the $c$-axis of the hexagonal magnet MnNiGa (Figure 2m).

The magnetic scattering cross section for SANS experiments can be written as:[29]

$$\sigma(\mathbf{q}) \propto \mathbf{M}_\perp(\mathbf{q}) \cdot \mathbf{M}_\perp^*(\mathbf{q}) \qquad (3)$$

where $\mathbf{M}_\perp$ is the magnetic structure factor that is projected onto the plane perpendicular to the scattering wavevector $\mathbf{q}$, and $\mathbf{M}_\perp^*$ is the complex conjugate of $\mathbf{M}_\perp$. The SANS pattern can thus be simulated by inserting different magnetic structures into Equation (3). At zero field, the bulk ground state is expected to be the stripe domain phase, or a labyrinth domain state.[4,19] Thus, in reciprocal space, the magnetic scattering lies on a sphere of radius of $q_m$ as shown by the yellow sphere in **Figure 2b**, where $q_m = 2\pi/L_D$, and $L_D$ is the periodicity of the stripe modulation. The simulated SANS pattern in this case is shown in **Figure 2c**, from which a ring-like pattern is expected, reflecting the random orientation and populations of the stripe domains. The scattering observed in the corresponding experimental data shown in **Figure 2d** is in agreement with the simulation. The radius of the ring is about 0.002 Å$^{-1}$, corresponding to the real-space stripe domain pitch $L_D$ of 300 nm, while our previous LTEM results showed this value from 180 to 230 nm.[19–21]

**Figure 2e** shows the $\mathbf{k}_i \perp \mathbf{B}$ geometry. Upon applying a magnetic field, three possible magnetic phases can evolve. The first one is the biskyrmion phase, as observed in LTEM studies.[18–23] This state will only have magnetic correlations in the plane perpendicular to **B**, as illustrated in **Figure 2f**, giving rise to two spots with $\pm q_x$ in the SANS pattern. The second phase is the single-domain cone phase, for which the propagation wavevector is along **B**,[5,14] as shown in **Figure 2i**. Consequently, one would expect two peaks with $\pm q_y$ (Figure 2j). The third possible state is the field-polarized state, i.e., the ferromagnetically aligned state, which leads to a strong peak at the origin of reciprocal space. It is therefore possible to unambiguously distinguish between these three phases in the $\mathbf{k}_i \perp \mathbf{B}$ geometry.

In our experiments, when applying a finite field, the ring-like pattern gradually transforms into a twofold-symmetric scattering pattern, as shown in **Figure 2h**. The magnetic correlation length ($2\pi/(0.0016$ Å$^{-1}) \approx 390$ nm) is comparable to that in our previous LTEM measured biskyrmion results (from 300 to 500 nm).[19] As discussed above, this is a clear indication that the stripe domains continuously evolve into the biskyrmion state. As SANS is a bulk probe, being relatively insensitive to surface effects, it can be concluded that biskyrmions stack along the *y*-direction, probably forming a tube-like three-dimensional structure (Figure 2e) analogous to regular skyrmions.[28] By further increasing the field from 0.38 to 0.5 T, the twofold-symmetric spots become weaker (Figure 2l), suggesting that biskyrmions become more weakly correlated and/or exist at lower density in this phase. Moreover, the two spots move closer to each other, which means that the inter-biskyrmion spacing increases with increasing field. The biskyrmion total scattering intensity, as well as the absolute value of the biskyrmion scattering wavevector as a function of magnetic field, is shown in **Figure 2p**. It is worth noting that throughout the phase diagram, we do not observe a conical phase.

Next, we investigate the $\mathbf{k}_i \parallel \mathbf{B} \parallel [001]$ geometry for a single-crystalline sample, which reveals the structure within the biskyrmion plane. When studying chiral skyrmions in helimagnets using the same geometry, the skyrmions form a long-range ordered hexagonal lattice, leading to six-fold-symmetric SANS peaks.[5] In SANS, the magnetic scattering factor $F_m(\mathbf{q}) = \mathbf{f}_{\text{motif}}(\mathbf{q}) * \mathbf{f}_{\text{lattice}}(\mathbf{q})$ is the convolution of the form factor of the biskyrmion motif, $f_{\text{motif}}(\mathbf{q})$, and their lattice order $f_{\text{lattice}}(\mathbf{q})$.[29] If the biskyrmions are not forming a periodic lattice, e.g., as in the disordered skyrmion phase hosted in the geometrically frustrated spin liquid material $Co_7Zn_7Mn_6$,[11] $f_{\text{lattice}}(\mathbf{q})$ smears out into an isotropic distribution in the $q_x$-$q_y$ plane, while the contribution of $f_{\text{motif}}(\mathbf{q})$ to $F_m(\mathbf{q})$ becomes more pronounced. Therefore, the non-lattice state offers a unique opportunity to study the form factor of the biskyrmion motif.[30]

**Figure 2o** shows the experimental biskyrmion SANS pattern obtained at 0.4 T. First, no diffraction peaks are observed, suggesting that in MnNiGa bulk crystals the biskyrmions are not long-range ordered in the hexagonal plane. Note that in LTEM imaging, owing to the small field-of-view, biskyrmions may form a distorted hexagonal lattice,[18–23] which is not in contradiction to the SANS data. In such a disordered state, a possible scenario is that individual biskyrmions have a random in-plane direction due to spontaneous symmetry breaking. Consequently, there would be no preferred direction of the biskyrmions when averaging over all quasiparticles. In this scenario, the scattering intensity is equally distributed forming a ring.

Interestingly, the experimental scattering pattern (Figure 2o) is anisotropic and has the shape of an astroid (a hypocycloid with four cusps). Therefore, it can be concluded that even though the biskyrmions are not long-range ordered, surprisingly their in-plane directions are aligned, and locked along certain directions, consistent with the LTEM measured results under small field-of-view.[19–21] As we will show the particular anisotropy in the scattering pattern directly reflects the form factor of the biskyrmion motif.

**Figure 3a-c** shows the experimental SANS patterns for different magnetic fields in the $\mathbf{k}_i \parallel \mathbf{B}$ geometry. Below 0.4 T, where the biskyrmion phase forms, the astroid shape remains almost unchanged, while the intensity varies with applied field. Most importantly, the fourfold-symmetric SANS pattern remains always locked along a direction, such that two of the four corners are along the crystallographic [100] directions, as shown in **Figure 3a** and **3b**. This finding points directly towards an intimate relationship between the in-plane biskyrmion direction and the magnetocrystalline anisotropy. The remanent field of the superconducting magnet, which is present during cool-down of the sample, may have a similar effect on the magnetic state as the field-cooling scenario reported by Peng *et al.*,[20] thereby explaining the ~0.03 T data. **Figure 3d** shows the magnetic field dependence of the magnetization measured under H ∥ [001]. Further increasing the field to saturation at 1.1 T leads to a smearing out of the fourfold-symmetric pattern, suggesting the transformation into the field-polarized state. The single-crystalline sample shows two easy-magnetization axes which are separated by 90° with respect to each other within the *ab*-plane, as obtained by the magnetic measurements (Figure 3e). We compared the results for the polycrystalline and the single-crystalline sample by plotting the reduced 1-dimensional data as shown in supporting information **Figure S1**. The main differences between field-polarized state and biskyrmion state appear from Q ≈ 0.001 to 0.01 Å$^{-1}$ for both data.

In order to quantitatively study the biskyrmion motif structure, numerical simulations of the SANS intensity were performed by inserting Equation (1) into Equation (3), and by using various combinations of structural biskyrmion parameters. By fitting the simulated form factors to the observed astroid-shaped experimental pattern, the bond distance *d*, shape factor *η*, as well as the in-plane biskyrmion direction is obtained for the different biskyrmion configurations. **Figure 4a** shows the form factor of a typical biskyrmion with *d* = 50 nm, *η* = 2, and Ψ = 185°, representing a biskyrmion non-lattice state (shown as inset, where brown arrows represent the direction of individual biskyrmions). It is clear that the form factor of the motif exhibits twofold symmetry, while the symmetry axis of the SANS pattern coincides with the $C_2$ axis of the biskyrmion configuration in real space.

However, there are more ways in which the alignment direction of the biskyrmions can be affected by the symmetry of the host crystal. The first is the threefold degeneracy of the biskyrmion order, as there are three equivalent [100] crystalline axes within the *ab*-plane. Consequently, a sixfold-symmetric form factor is expected, as shown in **Figure 4c**, which does not agree with the experimental data. Therefore, the only crystalline order that is compatible with the astroid-shaped diffuse scattering pattern is the formation of 90° separated two easy magnetization axes. The biskyrmion direction is locked along the easy-magnetization axis direction, giving rise to two possible directionally ordered states that are separated by 90°. As a result, the overall form factor appears as a fourfold-symmetric pattern, as shown in **Figure 4d**, and consistent with the in-plane magnetic anisotropy shown in **Figure 3e**. However, such a fourfold-symmetric pattern does not have the shape of an astroid. The parameter that changes the shape of the boundary from convex to concave is the bond distance *d*, as it is the key parameter that governs the detailed in-plane anisotropy described by the form factor. As shown in **Figure 4e**, by using an appropriate *d*, the simulated contrast fits the experimentally obtained pattern well. Lastly, we show that $\Psi$, the in-plane direction of the biskyrmions, directly correlates with the azimuthal rotation of the SANS pattern, see **Figure 4e** and **4f**.

Next, we discuss the possible energy terms that could govern such an 'order within disorder' phenomenon in MnNiGa bulk crystals. Here, order within disorder relates to the fact that despite the lack of positional order of the biskyrmion lattice, there is directional order, i.e., the biskyrmion axes all point in the same direction. Indeed, the formation of biskyrmions has not been accurately captured in the framework of micromagnetism, indicating that such quasiparticles may exist as a metastable state. A simple modification of the model is to introduce trigonal magnetocrystalline anisotropy $\omega_{cry}$ into Equation (2), which breaks the $O(2)$ symmetry. However, such a term should also have an effect on the anisotropy of the stripe domain as well. In our SANS experiments, the stripe domain populates random orientations, as evidenced by the ring-like scattering pattern, excluding the possibility that the magnetocrystalline energy is a first-order term. Therefore, the magnetocrystalline anisotropy does not play an important role in determining their positional arrangement, which is different from the case of $Cr_{11}Ge_{19}$.[22] Nevertheless, as the biskyrmions form at elevated fields, their polar axes may be affected by higher-order terms from the trigonal magnetocrystalline anisotropy, such that the $C_2$ axis is weakly locked along [100] directions with threefold degeneracy. The next term to consider is the biskyrmion-biskyrmion interaction, which is dominated by the effective force $\mathbf{F}(\mathbf{R}_i - \mathbf{R}_j) = -\nabla V$, where $\mathbf{R}_{i\,(j)}$ are the neighboring biskyrmion positions, and $V$ is the effective potential energy that results from Equation (2). It is thus clear that $V(\mathbf{r})$ is highly anisotropic due to the $C_{2v}$-shaped biskyrmion configuration. Therefore, the threefold degenerate directional order would tend to lock in one of the three [100] directions to minimize the total energy of the system. Moreover, as $\mathbf{F}(\mathbf{R}_i - \mathbf{R}_j)$ will also not be strong, since the biskyrmions do not display long-range order and form a lattice state.

In summary, we have determined the structural parameters of the biskyrmion state in MnNiGa bulk crystals. First, biskyrmions do not assemble into a long-range-ordered lattice. Instead, their correlation length always remains comparable to the periodicity of the stripe domains, and increases with increasing magnetic field. Second, although biskyrmions do not order into periodic lattices on a large length scale, they are in-plane rotationally ordered, with the rotation direction being locked to certain crystallographic directions. These polar molecular properties of the biskyrmion system suggest an intricate energy balance in MnNiGa, which is fundamentally different from chiral skyrmion systems. Such anisotropic nature, representing an extra degree of freedom for this type of skyrmions, may be exploited as information carrier. In this case, the two binary states are encoded within the same topological state, which reduces the energy consumption that is required for the state switching, demonstrating advanced device scheme for skyrmion-based memories.

**Experimental Section**

Sample preparation: Polycrystalline MnNiGa samples were synthesized by the growth method described in detail in reference [19], and single-crystalline samples were grown using the optical floating zone technique, using a *c*-axis oriented seed crystal. The growth direction is along the *c*-axis direction. The polycrystalline sample used for SANS measurements is a disc with a diameter of 13 mm and thickness of 1 mm. The single-crystalline sample used for SANS measurements is a cube measuring 2.5 × 2.5 × 2.5 mm$^3$, which was oriented using a Laue and an x-ray diffractometer. The data is shown in the main text in **Figure 1h**. The Curie temperature of the MnNiGa sample is ~350 K (as reported in reference [19]).

Magnetic measurements: Single-crystalline sample magnetization data were measured by applying a magnetic field of 0.3 T at 215 K in a Quantum Design Magnetic Property Measurement System (MPMS). The applied magnetic field is perpendicular to the *c*-axis and the sample is rotated 360° around the *c*-axis in increments of 1°. The data is shown in the main text in **Figure 3e**. The magnetic field dependence of the magnetization measured for H ∥ [001] at 180, 200, 220, and 250 K is shown in the main text in **Figure 3d**. The characteristic critical fields were measured by superconducting quantum interference device (SQUID) magnetometry for the polycrystalline sample, and shown to be qualitatively consistent those reported in our previous LTEM study.[19] Note that the critical fields defining the phase boundaries differ between thin film lamella and bulk samples due to demagnetization effects.

Neutron diffraction measurements: Neutron powder-diffraction measurements were performed at 400 K on the high-resolution powder diffractometer BT1 at the National Institute of Standards and Technology (NIST), USA. A Cu(311) monochromator was used to produce a monochromatic neutron beam of wavelength 1.5397 Å. The program FullProf was used for the Rietveld refinement of the crystal structures of the compound. The crystallographic structure of MnNiGa has a layered $Ni_2In$-type centrosymmetric hexangular structure with space group *P*63/*mmc*. The data is shown in the main text in **Figure 1j**. Other 300 K powder neutron diffraction patterns were obtained from the general purpose powder diffrctometer (GPPD)[31] (90° bank) at the China Spallation Neutron Source (CSNS) in Dongguan, China. Compared with the paramagnetic neutron diffraction data (Figure S2), we found no magnetic satellites in the 300 K data which would be characteristic for helical and/or conical magnetic structures. The reason for the absence of the helical magnetic peaks is that the helical periodicity is too large to be observed by neutron diffraction.

SANS measurements: SANS measurements on polycrystalline samples were performed using the SANS2d instrument[32,33] at the ISIS Neutron and Muon Source, Didcot, United Kingdom. The neutron beam was collimated over a length of 12 m before reaching the sample, and the scattered neutrons were collected by a 2-dimensional, position-sensitive multidetector placed 12 m behind the sample. The applied magnetic field is perpendicular to the neutron beam direction. The sample is measured at 215 K, which is the temperature at which the maximum topological Hall effect was measured[19] by neutrons with wavelengths from 1.75 Å to 12.5 Å with beam stop. The data reduction and analysis was done using the software package MantidPlot.[34] The field-polarized state data has been used as a background. The data are shown in the main text in **Figure 2d**, **2h**, **2l**, and **2p**. SANS measurements on the single-crystalline sample were performed using the D11 instrument[35] at the Institut Laue-Langevin (ILL), Grenoble, France. The neutron beam was collimated over a length of 37 m before reaching the sample, and the scattered neutrons were collected by a 2-dimensional, position-sensitive multi-detector placed 39 m behind the sample. The neutron wavelength was selected to be 8 Å. The circular aperture at the sample had a diameter of 7 mm. The sample is placed in a Gd jacket and coated with a layer of $Gd_2O_3$ to reduce the reflection by the surface of the sample. The applied magnetic field was parallel to the neutron beam and approximately parallel to the sample [001] axis. The *y*-axis rocking scans covered a range of up to ±10° with a SANS measurement performed every 1° and the *x*-axis rocking scans covered a range of up to ±3° with a SANS measurement performed every 1°, without beam stop (for an illustration of the measurement geometry see Figure 2a). The single-crystalline sample SANS patterns shown in the main text were obtained by summing over all rocking scans. The sample was cooled to 215 K in zero field and the rocking scans were subsequently measured at different applied magnetic fields, respectively. The data reduction and analysis was done using the software package GRASP.[36] The data are shown in the main text in **Figure 2o** and **Figure 3a**, **3b**, and **3c**. Initial single-crystalline sample SANS measurements were perforemd on SANSI at the Swiss Spallation Neutron Source, Paul Scherrer Institute (PSI), Villigen, Switzerland. The in-plane symmetry axis of the disordered biskyrmion system is not quite dependent on the temperature and sample magnetic field history. We compared the results for the polycrystalline and the single-crystalline sample in the reduced 1-dimensional data plot shown in **Figure S1**. There is more scattering intensity for the low field biskyrmion state compared to the high field polarized state in the *Q* region from 0.001 to 0.02 Å$^{-1}$ for the polycrystalline sample, and from 0.0008 to 0.01 Å$^{-1}$ for the single-crystalline sample.

**Supporting Information**
Supporting Information is available from the Wiley Online Library or from the author.

**Acknowledgements**
X. Y. L. and F. W. W. acknowledge financial support by the National Natural Science Foundation of China (No. 11675255) and the National Key R&D Program by MOST of China (No. 2016YFA0401503). S. L. Z. and T. H. acknowledge financial support by EPSRC (EP/N032128/1). W. H. W. acknowledge financial support by the National Key R&D Program by MOST of China (No. 2017YFA0303202) and the Key Research Program of the Chinese Academy of Sciences, KJZD-SW-M01. J. S. W. acknowledge financial support by the Swiss National Science Foundation (SNSF) via the Sinergia network "NanoSkyrmionics" (Grant No. CRSII5-171003). The neutron scattering experiments were carried out at the ISIS/STFC facility, which is sponsored by the Newton-China fund (proposal RB1620128). This work is based partly on experiments performed at the Swiss spallation neutron source SINQ, Paul Scherrer Institute, Villigen, Switzerland.

Received: ((will be filled in by the editorial staff))
Revised: ((will be filled in by the editorial staff))
Published online: ((will be filled in by the editorial staff))

References

[1]  L. D. Landau, *Zh. Eks. Teor. Fiz.* **1937**, *7*, 19.

[2]  N. D. Mermin, *Rev. Mod. Phys.* **1979**, *51*, 591.

[3]  B. Lebech, J. Bernhard, T. Freltoft, *J. Phys. Condens. Matter* **1989**, *1*, 6105.

[4]  N. Nagaosa, Y. Tokura, *Nat. Nanotechnol.* **2013**, *8*, 899.

[5]  S. Muhlbauer, B. Binz, F. Jonietz, C. Pfleiderer, A. Rosch, A. Neubauer, R. Georgii, P. Boni, *Science (80-. ).* **2009**, *323*, 915.

[6]  X. Z. Yu, Y. Onose, N. Kanazawa, J. H. Park, J. H. Han, Y. Matsui, N. Nagaosa, Y. Tokura, *Nature* **2010**, *465*, 901.

[7]  S. Heinze, K. von Bergmann, M. Menzel, J. Brede, A. Kubetzka, R. Wiesendanger, G. Bihlmayer, S. Blügel, *Nat. Phys.* **2011**, *7*, 713.

[8]  W. Jiang, P. Upadhyaya, W. Zhang, G. Yu, M. B. Jungfleisch, F. Y. Fradin, J. E. Pearson, Y. Tserkovnyak, K. L. Wang, O. Heinonen, S. G. E. te Velthuis, A. Hoffmann, *Science* **2015**, *349*, 283.

[9]  C. Moreau-Luchaire, C. Moutafis, N. Reyren, J. Sampaio, C. A. F. Vaz, N. Van Horne, K. Bouzehouane, K. Garcia, C. Deranlot, P. Warnicke, P. Wohlhüter, J.-M. George, M. Weigand, J. Raabe, V. Cros, A. Fert, *Nat. Nanotechnol.* **2016**, *11*, 444.

[10] Z. Hou, W. Ren, B. Ding, G. Xu, Y. Wang, B. Yang, Q. Zhang, Y. Zhang, E. Liu, F. Xu, W. Wang, G. Wu, X. Zhang, B. Shen, Z. Zhang, *Adv. Mater.* **2017**, *29*, 1701144.

[11] K. Karube, J. S. White, D. Morikawa, C. D. Dewhurst, R. Cubitt, A. Kikkawa, X. Yu,

Y. Tokunaga, T. H. Arima, H. M. Rønnow, Y. Tokura, Y. Taguchi, *Sci. Adv.* **2018**, *4*, eaar7043.

[12] I. Kézsmárki, S. Bordács, P. Milde, E. Neuber, L. M. Eng, J. S. White, H. M. Rønnow, C. D. Dewhurst, M. Mochizuki, K. Yanai, H. Nakamura, D. Ehlers, V. Tsurkan, A. Loidl, *Nat. Mater.* **2015**, *14*, 1116.

[13] A. O. Leonov, T. L. Monchesky, N. Romming, A. Kubetzka, A. N. Bogdanov, R. Wiesendanger, *New J. Phys.* **2016**, *18*, 065003.

[14] U. K. Rößler, A. A. Leonov, A. N. Bogdanov, *J. Phys. Conf. Ser.* **2011**, *303*, 012105.

[15] T. Adams, S. Mühlbauer, C. Pfleiderer, F. Jonietz, A. Bauer, A. Neubauer, R. Georgii, P. Böni, U. Keiderling, K. Everschor, M. Garst, A. Rosch, *Phys. Rev. Lett.* **2011**, *107*, 217206.

[16] S. Pöllath, J. Wild, L. Heinen, T. N. G. Meier, M. Kronseder, L. Tutsch, A. Bauer, H. Berger, C. Pfleiderer, J. Zweck, A. Rosch, C. H. Back, *Phys. Rev. Lett.* **2017**, *118*, 207205.

[17] A. K. Nayak, V. Kumar, T. Ma, P. Werner, E. Pippel, R. Sahoo, F. Damay, U. K. Rößler, C. Felser, S. S. P. Parkin, *Nature* **2017**, *548*, 561.

[18] X. Z. Yu, Y. Tokunaga, Y. Kaneko, W. Z. Zhang, K. Kimoto, Y. Matsui, Y. Taguchi, Y. Tokura, *Nat. Commun.* **2014**, *5*, 3198.

[19] W. Wang, Y. Zhang, G. Xu, L. Peng, B. Ding, Y. Wang, Z. Hou, X. Li, E. Liu, S. Wang, J. Cai, F. Wang, J. Li, F. Hu, G. Wu, B. Shen, X. X. Zhang, *Adv. Mater.* **2016**, *28*, 6887.

[20] L. Peng, Y. Zhang, W. Wang, M. He, L. Li, B. Ding, J. Li, Y. Sun, X.-G. Zhang, J. Cai, S. Wang, G. Wu, B. Shen, *Nano Lett.* **2017**, *17*, 7075.

[21] L. Peng, Y. Zhang, M. He, B. Ding, W. Wang, H. Tian, J. Li, S. Wang, J. Cai, G. Wu, J. P. Liu, M. J. Kramer, B. Shen, *npj Quantum Mater.* **2017**, *2*, 30.

[22] R. Takagi, X. Z. Yu, J. S. White, K. Shibata, Y. Kaneko, G. Tatara, H. M. Rønnow, Y.

Tokura, S. Seki, *Phys. Rev. Lett.* **2018**, *120*, 037203.

[23] J. C. T. Lee, J. J. Chess, S. A. Montoya, X. Shi, N. Tamura, S. K. Mishra, P. Fischer, B. J. McMorran, S. K. Sinha, E. E. Fullerton, S. D. Kevan, S. Roy, *Appl. Phys. Lett.* **2016**, *109*, 022402.

[24] S. Zhang, A. A. Baker, S. Komineas, T. Hesjedal, *Sci. Rep.* **2015**, *5*, 15773.

[25] E. A. Giess, *Science* **1980**, *208*, 938.

[26] F. Zheng, F. N. Rybakov, A. B. Borisov, D. Song, S. Wang, Z.-A. Li, H. Du, N. S. Kiselev, J. Caron, A. Kovács, M. Tian, Y. Zhang, S. Blügel, R. E. Dunin-Borkowski, *Nat. Nanotechnol.* **2018**, *13*, 451.

[27] F. N. Rybakov, A. B. Borisov, S. Blügel, N. S. Kiselev, *Phys. Rev. Lett.* **2015**, *115*, 117201.

[28] H. S. Park, X. Yu, S. Aizawa, T. Tanigaki, T. Akashi, Y. Takahashi, T. Matsuda, N. Kanazawa, Y. Onose, D. Shindo, A. Tonomura, Y. Tokura, *Nat. Nanotechnol.* **2014**, *9*, 337.

[29] B. T. M. W. and C. J. Carlile, *Experimental Neutron Scattering*, Oxford University Press, UK **2013**.

[30] J. S. White, E. M. Forgan, M. Laver, P. S. Häfliger, R. Khasanov, R. Cubitt, C. D. Dewhurst, M.-S. Park, D.-J. Jang, H.-G. Lee, S.-I. Lee, *J. Phys. Condens. Matter* **2008**, *20*, 104237.

[31] J. Chen, L. Kang, H. Lu, P. Luo, F. Wang, L. He, *Phys. B Condens. Matter* **2017**, *551*, 370.

[32] R. K. Heenan, S. M. King, D. S. Turner, J. R. Treadgold, *17th Meet. Int. Collab. Adv. Neutron Sources* **2005**.

[33] X. Y. Li, L. H. He, and F. W. Wang, *Dynamical Process of Helical-Biskyrmion Transformation of Hexagonal Magnet MnNiGa Probed by SANS*, **2016**, ISIS DOI: 10.5286/ISIS.E.82354759.

[34] O. Arnold, J. C. Bilheux, J. M. Borreguero, A. Buts, S. I. Campbell, L. Chapon, M. Doucet, N. Draper, R. Ferraz Leal, M. A. Gigg, V. E. Lynch, A. Markvardsen, D. J. Mikkelson, R. L. Mikkelson, R. Miller, K. Palmen, P. Parker, G. Passos, T. G. Perring, P. F. Peterson, S. Ren, M. A. Reuter, A. T. Savici, J. W. Taylor, R. J. Taylor, R. Tolchenov, W. Zhou, J. Zikovsky, *Nucl. Instruments Methods Phys. Res. Sect. A Accel. Spectrometers, Detect. Assoc. Equip.* **2014**, *764*, 156.

[35] X. Y. Li, R. Cubitt, and J. S. White, *Wide Temperature Range Biskyrmion Magnetic Nanodomains in MnNiGa Study by Small Angle Neutron Scattering*, **2018**, Institut Laue-Langevin (ILL) doi:10.5291/ILL-DATA.5-42-460.

[36] C. Dewhurst, GRASP, **2018**, https://www.ill.eu/en/users/support-labs-infrastructure/software-scientific-tools/grasp/.

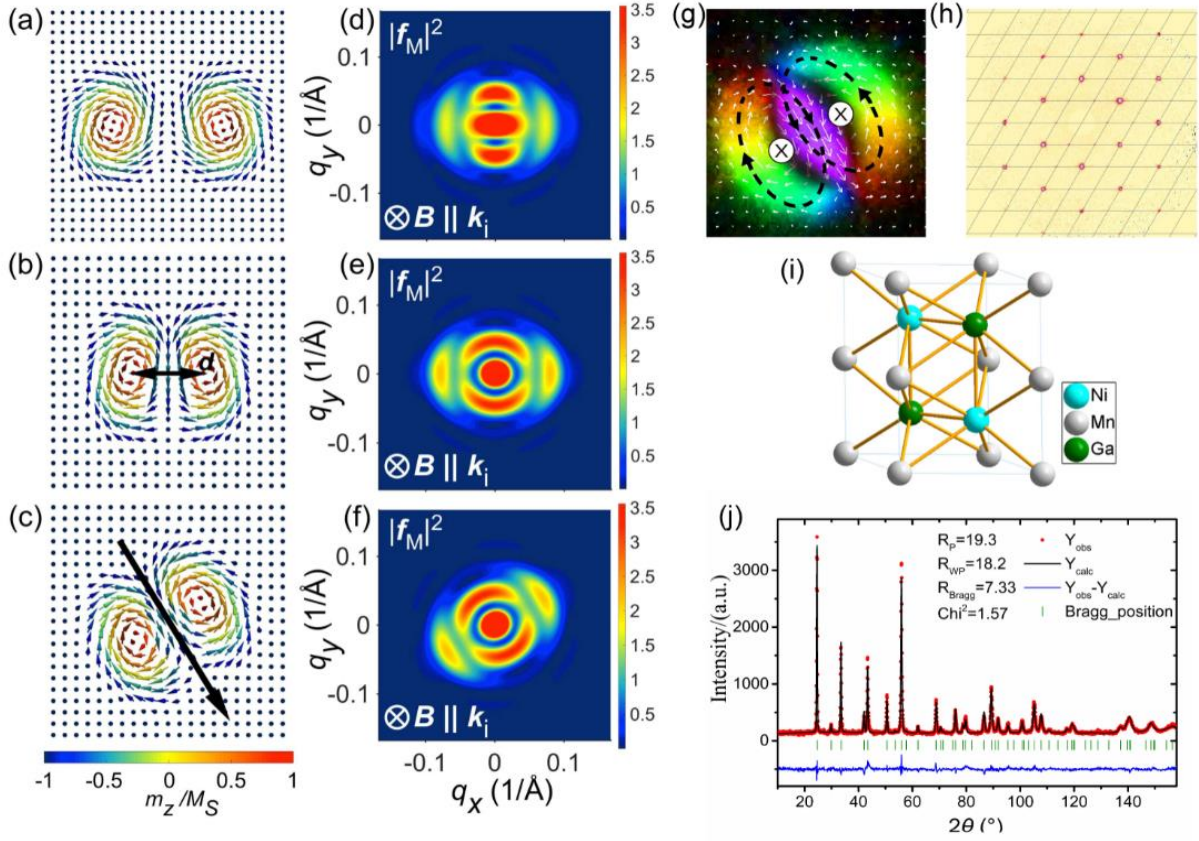

**Figure 1**. a-c) Structural parameters describing a $C_{2v}$ biskyrmion. The polarities are the same for all three plots, namely $\lambda = 1$. Two $D_n$ skyrmions are brought together in a ferromagnetic background with an inter-skyrmion (bond) distance $d$. Starting from the structure shown in a), $d$ is further decreased, while the biskyrmion shape deforms into an ellipse ($\eta = 2$). The principle symmetry axis is labeled by the black arrow, which gives also the direction of the biskyrmion ($\Psi = 30°$). d-f) The (a-c) corresponding simulated magnetic form factor (where $|f_M|^2$ is $|f_{motif}(\mathbf{q})|^2$), respectively. g) An in-plane magnetization of the biskyrmion configuration. h) The MnNiGa single-crystalline sample is checked by single-crystal x-ray diffractometer. The data shows the sample has hexagonal lattice structure. i,j) The refinement result of high-resolution neutron powder diffraction data which demonstrated the structure of MnNiGa is a centrosymmetric hexangular structure. The Mn atoms located at 2a site, Ni atoms located at 2d and Ga atoms located at 2c with a space group of *P*63/*mmc*.

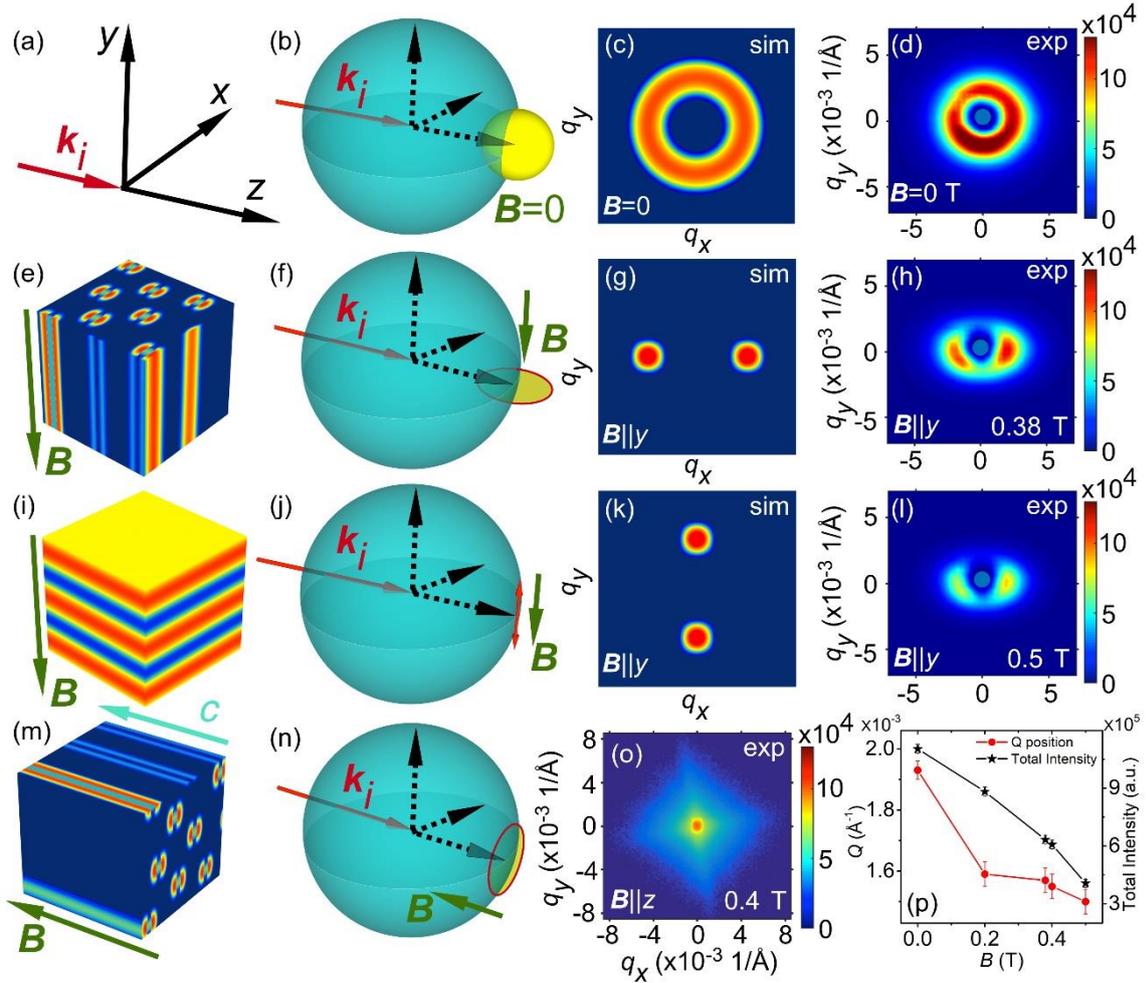

**Figure 2.** a) Coordinate system. b,f,j,n) Ewald sphere representations for the various experimental SANS configurations. c) Simulated and d) experimental SANS pattern in zero field showing a ring-like contrast. e) Illustration of the 3D magnetization distribution of the disordered biskyrmion phase in an applied out-of-plane field of 0.38 T. g,h) Simulated and experimental SANS pattern for the corresponding state. i) Assumed conical state in an applied out-of-plane field of 0.5 T, and k), corresponding simulated SANS contrast. l) The experimental SANS data do not support the existence of the conical state. m) Biskyrmion arrangement for $\mathbf{k}_i \parallel \mathbf{B}$, and o), corresponding experimental SANS pattern. p) The absolute value of the biskyrmion scattering wavevector, as well as the associated total scattering intensity as a function of the magnetic field for $\mathbf{k}_i \perp \mathbf{B}$ geometry. The panels shown in (d,h,l,p) are measured on the SANS2d instrument at ISIS for the polycrystalline sample, and the panel shown in (o) is measured on the D11 instrument at ILL for the single-crystalline sample (see experimental section for more details).

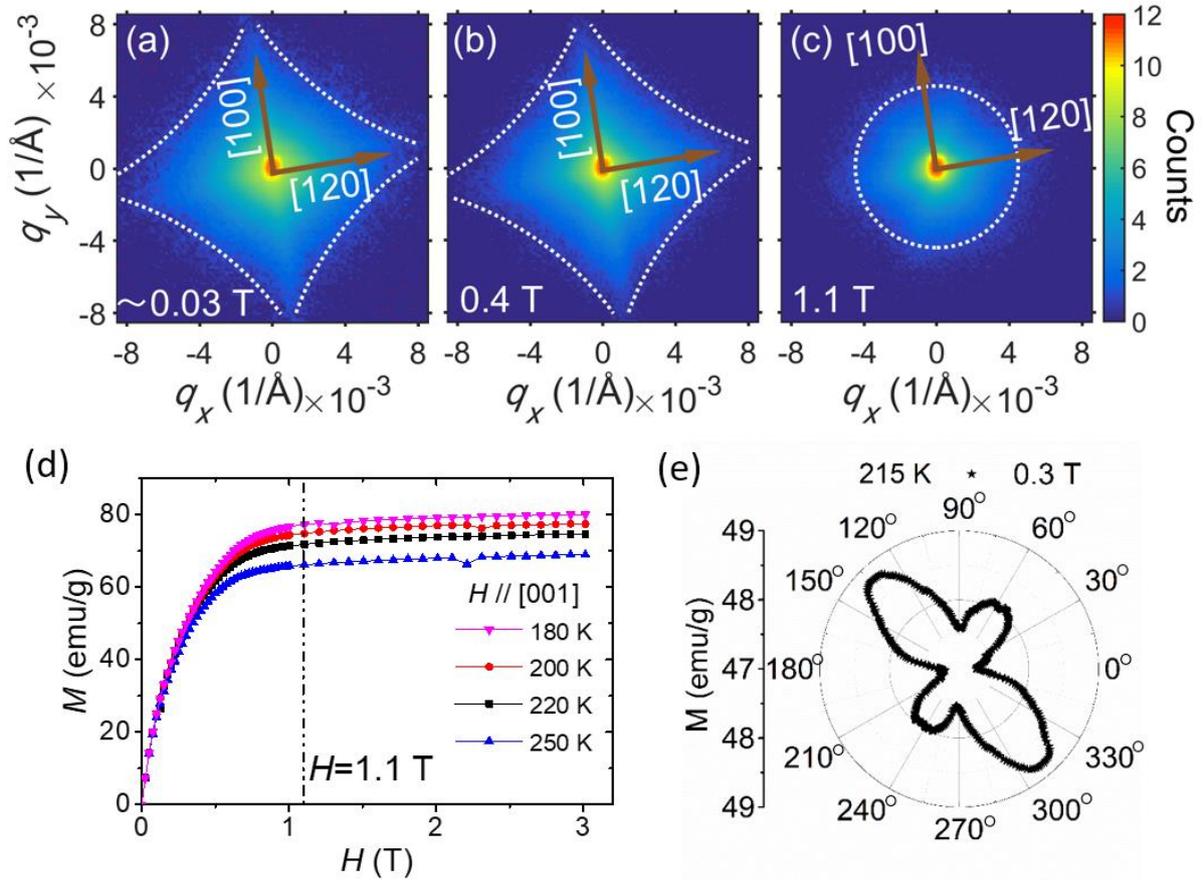

**Figure 3**. Evolution of the experimental SANS pattern in the $k_i \parallel B$ configuration for applied fields of a) ~0.03 T (remanent field of the superconducting magnet), b) 0.4 T, and c) 1.1 T. Note that the color scale is logarithmic to make weak features visible. d) Magnetic field dependence of the magnetization measured under H ∥ [001] at 180, 200, 220, and 250 K, respectively. e) The single-crystalline sample shows two easy-magnetization axes which are separated by 90° with respect to each other within the ab-plane as obtained by the magnetic measurements.

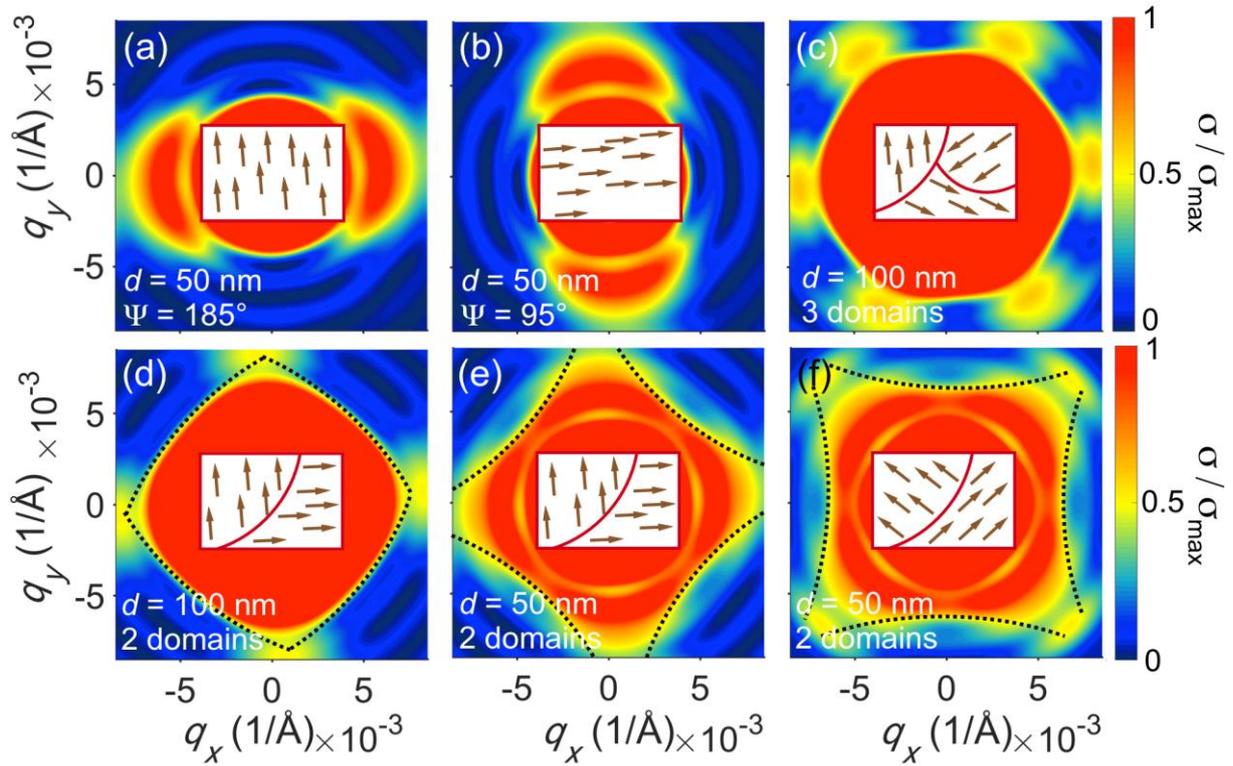

**Figure 4.** SANS contrast simulation for (laterally) disordered biskyrmion lattices. a) Single domain oriented at Ψ = 185°, and b) Ψ = 95°, and c) three domains separated by 120° in-plane rotation. Two domains (90° rotated) for d) $d$ = 100 nm and e) 50 nm. f) Same as in (e) with an in-plane rotation of ΔΨ = 45°. For all simulations, $\eta$ = 2.

**Oriented Three-Dimensional Magnetic Biskyrmion in MnNiGa Bulk Crystals**

*Xiyang Li, Shilei Zhang, Hang Li, Diego Alba Venero, Jonathan S White, Robert Cubitt, Qingzhen Huang, Jie Chen, Lunhua He, Gerrit van der Laan, Wenhong Wang, Thorsten Hesjedal, and Fangwei Wang\**

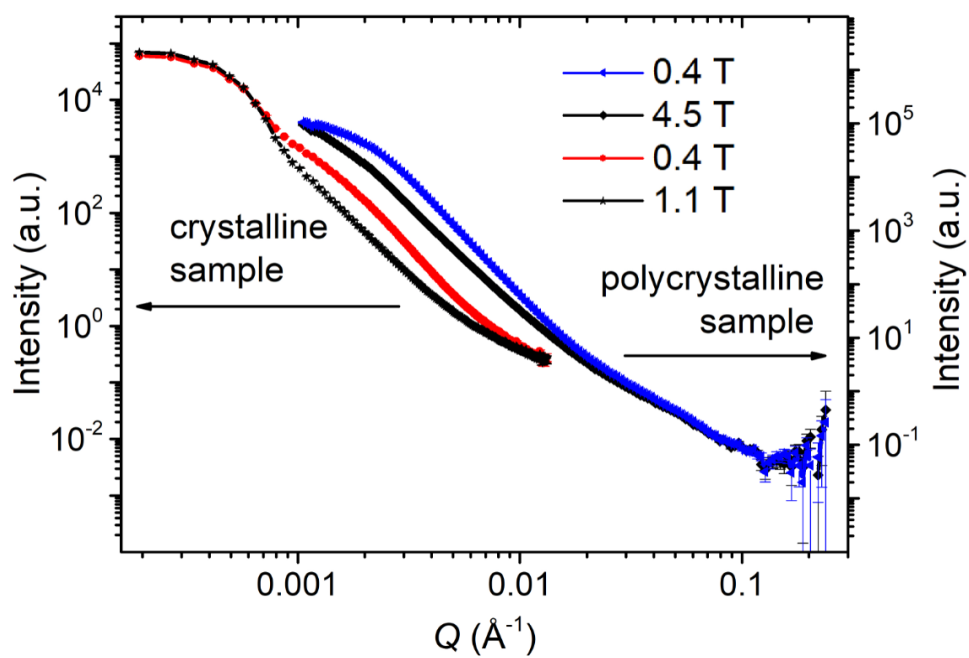

**Figure S1**: Reduced 1-dimensional data for the polycrystalline sample measured on SANS2d@ISIS and for the single-crystalline sample measured on D11@ILL. The scattering intensity for the low field biskyrmion state is larger compared to the high-field polarized state.

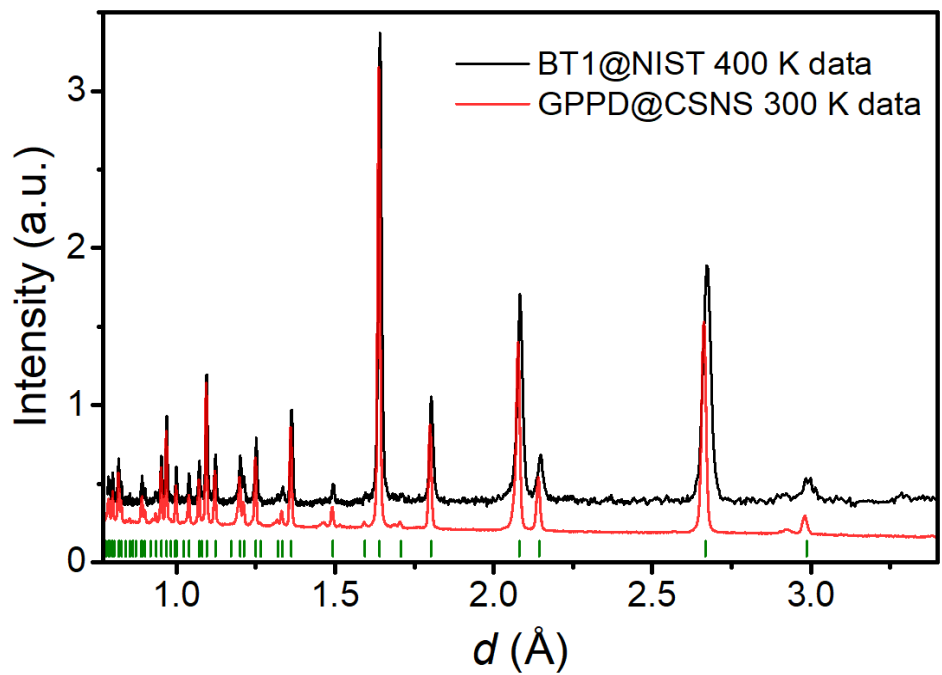

**Figure S2**: Comparison of neutron diffraction data obtained at 300 K and 400 K. The Curie temperature of the sample is ~350 K (as reported in reference [19]). Note that no magnetic satellites associated with the helical/conical phase are measured at 300 K as the helical period is too large to be detected by neutron diffraction. The green lines represent the Bragg peak positions of the crystalline structure.